\documentclass[reprint,superscriptaddress,amsmath,amssymb,aps]{revtex4-1}
\usepackage{}
\usepackage{cases}
\usepackage{amsmath}
\usepackage{amssymb}
\usepackage{amsfonts}
\usepackage{amssymb}
\usepackage{dcolumn}
\usepackage{bm}
\usepackage{hyperref}
\usepackage{graphicx}
\usepackage{upgreek}
\usepackage{subfigure}

\begin{document}
\title{NV-Center Based Digital Quantum Simulation of a Quantum Phase Transition in Topological Insulators}
\author{Chenyong Ju}
\affiliation{Hefei National Laboratory for Physics Sciences at the Microscale and Department of Modern Physics, University of Science
and Technology of China, Hefei, 230026, China}
\author{Chao Lei}
\affiliation{Hefei National Laboratory for Physics Sciences at the Microscale and Department of Modern Physics, University of Science
and Technology of China, Hefei, 230026, China}
\author{Xiangkun Xu}
\affiliation{Hefei National Laboratory for Physics Sciences at the Microscale and Department of Modern Physics, University of Science
and Technology of China, Hefei, 230026, China}
\author{Dimitrie Culcer}
\affiliation{School of Physics, The University of New South Wales, Sydney 2052, Australia}
\author{Zhenyu Zhang}
\affiliation{ICQD/HFNL, University of Science and Technology of China, Hefei, Anhui, 230026, China}
\author{Jiangfeng Du}\altaffiliation{Corresponding author: djf@ustc.edu.cn}
\affiliation{Hefei National Laboratory for Physics Sciences at the Microscale and Department of Modern Physics, University of Science
and Technology of China, Hefei, 230026, China}


\begin{abstract}

Nitrogen-vacancy centers in diamond are ideal platforms for quantum simulation, which allows one to handle problems that are intractable theoretically or experimentally. Here we propose a digital quantum simulation scheme to simulate the quantum phase transition occurring in an ultrathin topological insulator film placed in a parallel magnetic field [Zyuzin \textit{et al.}, Phys. Rev. B \textbf{83}, 245428 (2011)]. The quantum simulator employs high quality spin qubits achievable in nitrogen-vacancy centers and can be realized with existing technology. The problem can be mapped onto the Hamiltonian of two entangled qubits represented by the electron and nuclear spins. The simulation uses the Trotter algorithm, with an operation time of the order of 100 $\mu$s for each individual run.

\end{abstract}


\maketitle

The identification of a single nitrogen-vacancy (NV) color center \cite{Gruber97} has been a milestone in the development of diamond-based quantum technologies. Promising applications range from quantum information processing \cite{Fuchs09,Neumann10,van12,Shi10,Togan10,Bernien13} to various quantum metrology protocols, such as nanoscale magnetometry \cite{Balasubramanian08,Maze08,Staudacher13} and thermometry \cite{Kucsko13,Neumann13}. NV center spins are good candidates for these applications due to their easy initialization and readout using lasers \cite{Gruber97}, fast and high-performance spin manipulation \cite{Fuchs09,Xu12}, long coherence times at room temperature \cite{Balasubramanian09}, and the ability of accessing and manipulating nearby nuclear spins that facilitates a highly controllable few-spin system \cite{Neumann08,van12,Dutt07,Fuchs11}. Several promising schemes to scale the system to a larger number of entangled spins have also been proposed \cite{Yao12,Rabl10}. In recent years, NV-center samples of extraordinary quality have been synthesized \cite{Balasubramanian09}.


An important application of NV centers that may be realized in the near future is quantum simulation. Quantum simulators \cite{Lloyd96,Buluta09,Cai2013}, which serve as pure quantum systems, offer an exciting pathway towards problems that cannot be solved theoretically or tested experimentally. One area of great interest where theory and experiment are beginning to diverge is topological insulators (TI). These represent a new class of topological materials that are insulating in the bulk but conducting along the edges (2D TI) \cite{Kane05,Bernevig06} or surfaces (3D TI) \cite{Fu07}. The surface states of 3D TI are protected by time-reversal symmetry and describe massless Dirac fermions with spin-momentum locking \cite{Qi11}. A host of fascinating phenomena have been predicted, including charge-induced magnetic monopoles \cite{Qi09} and Majorana zero modes at the interface with a superconductor \cite{Fu08}, which can be used for topological quantum computation. Yet their experimental observation is hampered by a bottleneck in materials science. Current TI have low bulk band gaps and are sensitive to unintentional doping or defects, yielding a sizable bulk conductivity that hinders the observation of surface states in transport \cite{Hasan10}. An attractive method of dealing with this problem is to grow thin film TI \cite{Zhang10,Jiang12}, reducing the contribution of the bulk. Ultrathin TI films have interesting physical properties in their own right, including massive Dirac fermions \cite{Lu10}, nonlinear screening \cite{Liu13}, the magneto-optical effect \cite{Tse10}, excitonic superfluidity \cite{Seradjeh09},  quantum anomalous Hall effect \cite{Yu10} and a quantum phase transition (QPT) as a function of the in-plane magnetic field \cite{Zyuzin11}. However, producing high quality ultrathin film TI has been proven difficult with present technology \cite{Culcer12} (i.e., molecular beam epitaxy), with impurities and defects in the substrate and sample obscuring surface properties.

It has been proposed to use analog quantum simulators such as optical lattices to mimic TI \cite{Mazza12,Bermudez10,Goldman10}. In this paper, we propose a universal quantum simulation method to simulate a QPT in an ultrathin TI film in a parallel magnetic field, demonstrating the use of NV center spins to study the nontrivial physics emerging in these materials. This scheme, which could be construed as an artificial TI system, would constitute the first NV-center based quantum simulation. Tackling a nontrivial problem whose solution is known, it offers a reliable way to test the simulator. We anticipate further problems, including interacting TI, could be simulated and studied using NV center simulators in analogous architectures.

TI have one massless Dirac fermion state on each surface \cite{Liu10}. When the thickness of the bulk TI becomes small enough (often several quintuple layers) to be comparable to the \textit{penetration depth} of the surface states into the bulk, the two surfaces begin to hybridize \cite{Zyuzin11}. The Dirac fermion states acquire a mass, and the Hamiltonian of the thin film TI becomes $ H_{TI} = A (k_y \sigma _x - k_x \sigma _y)\, \tau _z + \Delta \tau _x$, where the material parameter $ A $ is the spin-orbit constant, $ \vec{k} $ is the electron wave vector and $ \vec{\sigma } $ is the vector of Pauli matrices. The vector $ \vec{\tau } $ represents the pseudospin Pauli matrices, which accounts for the presence of two surface layers, while $ \Delta $ is the tunnel coupling between the two layers, and is inversely proportional to the thickness of the bulk material.

\begin{figure}[tbp]
\centering
\includegraphics[width=\columnwidth]{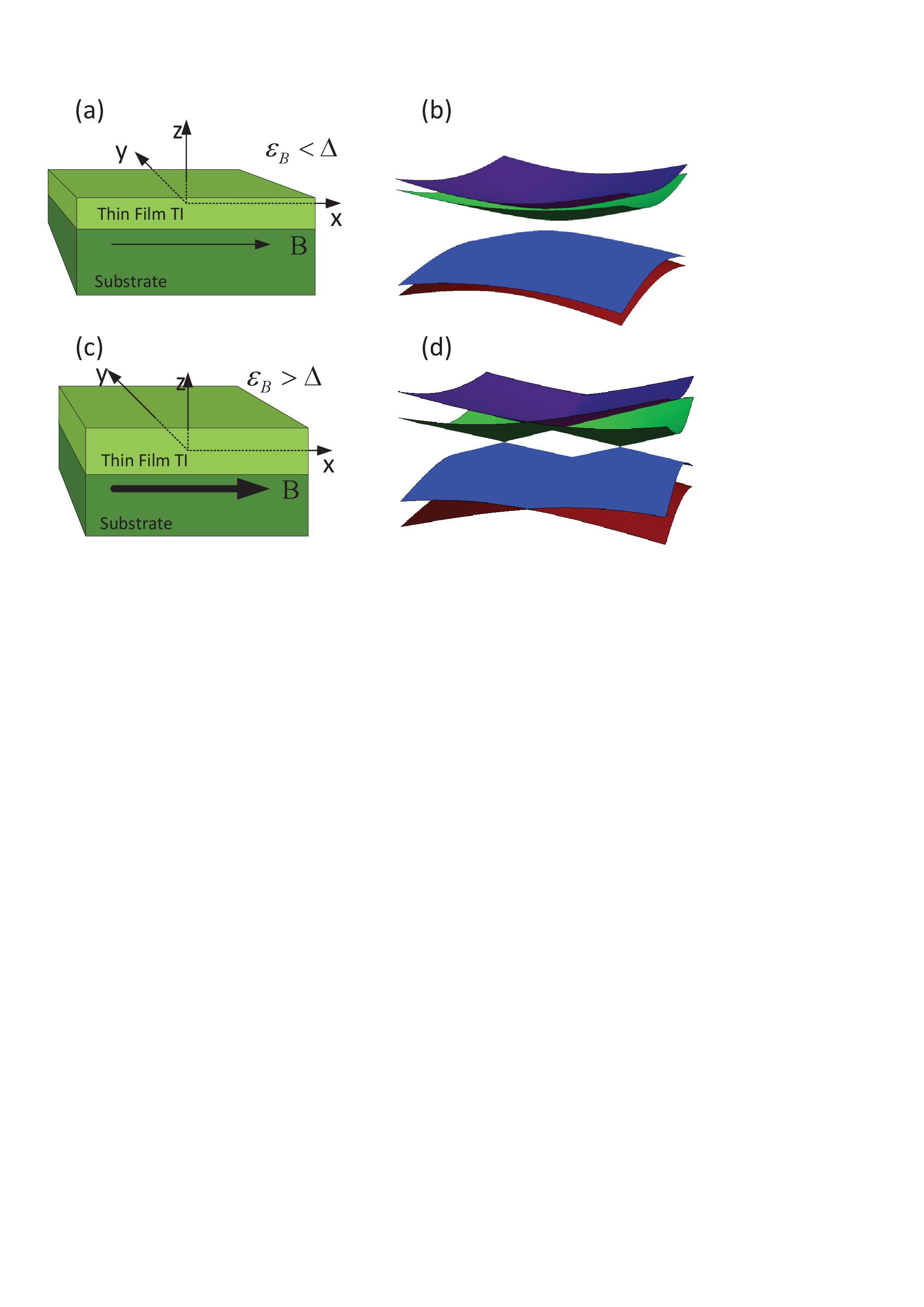}
  \caption{(Color online).
   (a) and (c) Ultrathin film TI in a parallel magnetic field, usually grown on a substrate. The magnetic field is oriented along the x-axis and the top and bottom surfaces are assumed to be in the planes $ z = \pm d $, where $ d $ is the thickness of the thin film TI (small enough that the two surfaces hybridize). In (a) $ \varepsilon _ B < \Delta $ and in (c) $\varepsilon_B > \Delta $.
   (b) and (d) The band structure of the surface states in a parallel magnetic field and the coupling of the two surfaces. When $ \varepsilon _ B < \Delta $ the thin film TI is gapped as shown in (b); as the magnetic field increases, there appear two Dirac cones as shown in (d).
}
  \label{fig1}
\end{figure}

It has been predicted theoretically that in ultrathin film TI a QPT will occur when an in-plane magnetic field is applied \cite{Zyuzin11}. Consider a magnetic field ${\bm B} \parallel \hat{x}$ and assume that the top and bottom surfaces of the thin film are located at $ z = \pm d/2 $ shown in Fig. \ref{fig1}. Then the Hamiltonian is \cite{Zyuzin11}
\begin{equation}\label{magnetic_Hamiltonian}
    {H_{TI}^B} = A({k_y}{\sigma _x} - {k_x}{\sigma _y}) \, {\tau _z} - {\varepsilon _B}{\sigma _x} + \Delta {\tau _x},
\end{equation}
where $ \varepsilon _B $ is the \textit{magnetic energy} with $ \varepsilon _B = v_F q_B $ \cite{Zyuzin11}, in which $ v_F $ is the Fermi velocity and $q_B = d/2l^2 -1/2mv_Fl^2 $ is the magnetic wave vector, with $ l = \sqrt{\hbar/eB} $ the magnetic length. The first term in $ k_B$ comes from the Aharonov-Bohm phase gradient and the second term comes from the Zeeman coupling. Tuning the magnetic field leads to a change in $\varepsilon _B$. The physics of the system changes drastically as a function of the parameter $\varepsilon_B/\Delta$. A QPT takes place at the value $ \varepsilon _B = \Delta$ from an insulating state with a diamagnetic response to a semimetallic state with zero magnetic response. The magnetic susceptibility plays the role of an order parameter. When $ \varepsilon _B < \Delta $ the thin film is in an insulating phase. As $\varepsilon _B $ is increased to a value $ \varepsilon _B > \Delta $, two Dirac cones appear, which can be characterized by a topological invariant, i.e. the spin winding number around any surface in momentum space \cite{Zyuzin11}. Consequently, the transition is topological: the two sides have different numbers of pairs of Dirac points. This field-driven semimetal-insulator transition does not occur for TI states on separated surfaces, but only when two surfaces are coupled.

\begin{figure}[tbp]
\centering
\includegraphics[width=0.8\columnwidth]{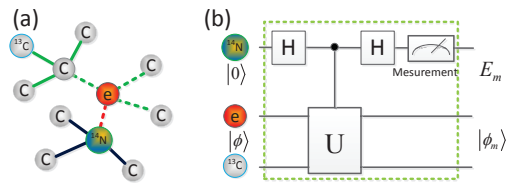}
  \caption{(Color online).
   (a) Nitrogen-Vacancy center in diamond.
   (b) The Quantum circuit describing the simulation (see the text for a detailed description).}
  \label{fig2}
\end{figure}

However, the calculated susceptibility is two orders of magnitude less than that of ordinary metals, and could be overwhelmed experimentally by the signal from the bulk TI. Moreover, given the difficulty in producing sufficiently clean TI samples, the QPT is difficult to observe in present-day TI. It is, however, possible to simulate the transition using a digital quantum simulator consisting of the spins of a NV center. An NV center is a defect in diamond consisting of a substitutional nitrogen atom and a vacancy at the nearest neighbor lattice position [Fig. \ref{fig2} (a)]. Two unpaired electrons on the center form an effective electron spin with a spin triplet ground state (S=1). The nuclear spin of the substitutional $^{14}$N atom and a a well-resolved $^{13}$C nuclear spin nearby are also employed in our simulation. The Hamiltonian of the three spins in an external magnetic field $B_0$ is ($\hbar=1$)
\begin{equation}\label{NV_Hamiltonian}
\begin{split}
  {H_0} = DS_z^2 + {\gamma _e}{B_0}{S_z} - {\gamma _C}{B_0}{C_z} - {\gamma _N}{B_0}{N_z}  \\
   + Q{N^2_z} + {J_C}{S_z}{C_z} + {J_N}{S_z}{N_z}.
\end{split}
\end{equation}
Here $S_z$, $C_z$ and $N_z$ are the spin operators of the electron spin (spin-1), the $^{14}$N nuclear spin (spin-1), and $^{13}$C nuclear spin (spin-1/2), respectively. $ D/2\pi = 2.87 $ GHz is the zero-splitting tensor of the electron spin, which originates from the the anisotropic magnetic dipole-dipole interaction between the two unpaired electrons. The second to fourth terms in the Hamiltonian are the Zeeman interactions of each spin with the gyromagnetic ratio $\gamma _e/2\pi = 2.8$ MHz/Gauss, $\gamma _C/2\pi = 1.1$ kHz/Gauss, $\gamma _N/2\pi = 0.3077 $ kHz/Gauss. $ Q/2\pi = -5.1 $ MHz is the quadrupole splitting tensor of the nitrogen nuclear spin. The hyperfine coupling constant between the electron spin and the $^{13}$C nuclear spin is selected to be $J_C/2\pi = 14 $ MHz \cite{Jiang09} to fulfill the parameter relations in this simulation. The hyperfine coupling constant between the electron spin and the $^{14}$N nuclear spin is $J_N/2\pi = 2.1 $ MHz. In the scheme only the $m_S = 0$ and $1$ levels of the electron spin are used, which can be treated as a spin-1/2 system. It is used to encode the $ \sigma $ pseudospin in the TI Hamiltonian (Eq. (\ref{magnetic_Hamiltonian})). The $^{13}$C nuclear spin is used to encode the $\tau$ pseudospin, and the $^{14}$N nuclear spin is the readout spin.

The simulation consists of a quantum algorithm of finding eigenvalues \cite{Abrams99}. The procedure is illustrated in Fig. \ref{fig2}(b). At the beginning, the $^{14}$N nuclear spin is prepared in $ \left| 0 \right\rangle $. The electron spin and $^{13}$C nuclear spin are prepared in an arbitrary state $ \left| \phi  \right\rangle = \sum\limits_m {{c_m}} \left| {{\phi _m}} \right\rangle $, where $ {\left| {{\phi _m}} \right\rangle} $ are the eigenstates of $ H_{TI}^B $, satisfying $ H_{TI}^B \left| {{\phi _m}} \right\rangle = E_m \left| {{\phi _m}} \right\rangle$. Such an initialization step can be conveniently realized in NV centers. The polarization of the NV electron spin and the nuclear spins has been demonstrated in Refs. \cite{Steiner10,Jiang09}. Appropriate microwave (MW) and radio-frequency (RF) pulses would prepare the electron spin and $^{13}$C nuclear spin into the state $ \left| \phi  \right\rangle$. After initialization, a Hadamard gate on the $^{14}$N nuclear spin transforms it into the superposition state $ (\left| 0 \right\rangle + \left| 1 \right\rangle)_N $. The simulation of the TI takes place in the $ \left| 1 \right\rangle _N $ subspace, represented by a $^{14}$N nuclear spin controlled-$U$ operation ($ U = e^{-i H_{TI}^B t} $) that transforms the spin state into
\begin{equation}
    \sum\limits_m {{c_m}} {\left| {{\phi _m}} \right\rangle} ({\left| 0 \right\rangle}_N + { e^{-i{E_m}t}}{\left| 1 \right\rangle}_N ).
\end{equation}
After another Hadamard gate (the reverse quantum Fourier transformation), the $^{14}$N nuclear spin is measured. By repeating the procedure with different evolution times $t$, each eigenenergy $E_m$ can be obtained with the weight $ |c_m|^2$. The measurement of the $^{14}$N nuclear spin can be achieved by swapping its state with the NV electron spin and then measure the electron spin.

\begin{figure}[tbp]
\centering
\includegraphics[width=1\columnwidth]{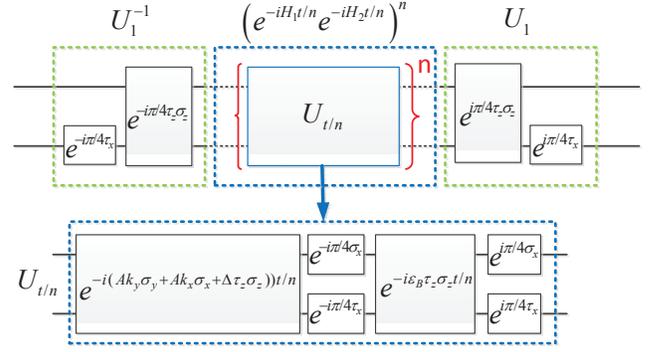}
  \caption{(Color online).
   Quantum circuit describing the time evolution $e^{-iH_{TI}^B t}$. The unitary transformation $ U_1 $ is realized in the green dotted rectangle frame and in the blue dotted frame the unitary operator $ e^{iH_{S}^Bt} $ is realized using the Trotter approximation. The bottom quantum circuit details that in the top dotted blue frame.
}
  \label{fig3}
\end{figure}

The controlled-$U$ operation plays the central role in the simulation. For simplicity, we first describe how to simulate $ H_{TI}^B $ with the electron spin and $^{13}$C nuclear spin under their hyperfine coupling ${J_C}{S_z}{C_z}$, and then illustrate how to realize it in the $ \left| 1 \right\rangle _N $ subspace, i.e., the controlled-$U$ operation. To simulate Eq. (\ref{magnetic_Hamiltonian})), the following transformation is used to map the TI Hamiltonian to an experimentally realizable Hamiltonian
\begin{align}
H_{S}^B  & = U_1^{-1} H_{TI}^B U_1\\
& = A{k_y}{\sigma _y} + A{k_x}{\sigma _x} + \Delta {\tau _z}{\sigma _z} + {\varepsilon _B}{\tau _y}{\sigma _y},\label{Hamiltonian_transform}
\end{align}
where $U_1 = {e^{i\pi /4{\tau _z}{\sigma _z}}} {e^{i\pi /4{\tau _x}}}$. The evolution of $e^{-iH_{S}^B t}$ could be further divided into two parts by employing the Trotter approximation method \cite{Lloyd96} as
\begin{equation}
    {e^{-iH_{S}^Bt}} \approx {\left( {{e^{ - i{H_1}t/n}}{e^{ - i{H_2}t/n}}} \right)^n},
\end{equation}
where $H_{S}^B = H_1 + H_2$, $ H_1 = {\varepsilon _B}{\tau _y}{\sigma _y} $, and $ H_2 = A{k_y}{\sigma _y} + A{k_x}{\sigma _x} + \Delta {\tau _z}{\sigma _z} $ (the whole quantum circuit is shown in Fig. \ref{fig3}). The time evolution operator $e^{-iH_{S}^Bt}$ can be realized within arbitrary accuracy by selecting an appropriate $n$ \cite{Lloyd96}, while $e^{-iH_1 t}$ can be realized by the free evolution under the hyperfine coupling ${J_C}{S_z}{C_z}$ and local transformation $U_2 = {e^{i\pi /4{\tau _x}}}{e^{i\pi /4{\sigma _x}}}$, where $ U_2 e^{-i {\varepsilon _B}{\tau _z}{\sigma _z} t} U_2^{-1} = e^{-iH_1 t} $. The Hamiltonian $H_2$ is exactly the Hamiltonian in the rotating frame when a microwave pulse $B_{1}\cos (\omega t+\phi )\mathbf{e}_{x}+B_{1}\sin (\omega t+\phi )\mathbf{e%
}_{y}$ is applied to the NV electron spin. The transformations $U_1$ and $U_2$ could also be conveniently realized in experiment, as $e^{i\pi /4{\tau _x}}$ and $e^{i\pi /4{\sigma _x}}$ represent the single-spin rotations of the $^{13}$C nuclear spin and NV electron spin respectively, and $e^{i\pi /4{\tau _z}{\sigma _z}}$ can be realized by a free evolution under the hyperfine coupling.

Here we have the parameter relations $Ak_x = \frac{1}{4} \gamma _e B_1 \cos\phi$, $Ak_y = \frac{1}{4} \gamma _e B_1 \sin\phi$, and $\Delta = \frac{1}{4} J_C$. To observe the phase transition, one needs to adjust the value of $\varepsilon _B$ to different domains $\varepsilon _B \leq \Delta$ and $\varepsilon _B > \Delta$. Letting $\varepsilon _B = s \Delta$, then $e^{ - i{H_1}{t}/n} = e^{ - i{\Delta}{\tau _y}{\sigma _y}st/n}$, i.e., by changing the evolution time of $H_1$ to $t^\prime = st$ the value of $\varepsilon _B$ could be efficiently adjusted.

The evolution $e^{-iH_{S}^Bt}$ should take place in the $ \left| 1 \right\rangle _N $ subspace. This can be achieved essentially due to the hyperfine coupling between the $^{14}$N nuclear spin and the NV electron spin, which makes the selective electron spin operations corresponding to different $^{14}$N nuclear spin states possible. The detailed method is as follows: (1) With selective microwave pulses, $e^{-i{H_2}t}$ can be realized in the $ \left| 1 \right\rangle _N $ subspace, while in the $ \left| 0 \right\rangle _N $ subspace selective electron spin $\pi $ pulses are performed to refocus the coupling between the electron spin and the $^{13} C$ nuclear spin and an effective identity operation is realized in this subspace; (2) By the same refocusing technique the free evolution in $ e^{-i{H _1 t}} $ can be realized only in the $ \left| 1 \right\rangle _N $ subspace; (3) $U_1$, $U_2$, and their inversion operations are performed in all subspaces, but since they occur in pairs a net identity operation is achieved in the $ \left| 0 \right\rangle _N $ subspace.

\begin{figure}[tbp]
\centering
\includegraphics[width=1\columnwidth]{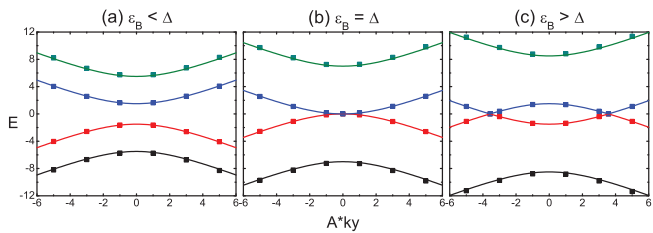}
  \caption{(Color online).
Numerical simulation results for the TI system eigenenergies, showing a QPT behavior. The curves are the theoretically predicted values. $ k_x = 0 $ is selected in the calculation since the QPT is illustrated most clearly in this case: (a) $\varepsilon _B / \Delta \approx 0.57 $, insulating phase with no Dirac points; (b) $\varepsilon _B / \Delta = 1 $, the QPT critical point; (c) $\varepsilon _B / \Delta \approx 1.43 $, semimetallic phase with 2 Dirac points.
}
  \label{fig4}
\end{figure}

We have performed a numerical simulation of the above control pulses acting on the NV-$^{13}$C-$^{14}$N spin system, and obtained the TI Hamiltonian (Eq. (\ref{magnetic_Hamiltonian})) eigenenergies according to the algorithm. The results faithfully reproduce the QPT behavior occurred in this TI system (Fig. \ref{fig4}): while increasing $\varepsilon _B$ from $\varepsilon _B < \Delta$ to $\varepsilon _B > \Delta$, the system changes from an insulating phase to a semimetallic phase. The results are agree with the theoretical prediction by Zyuzin et al. \cite{Zyuzin11}.

In practice, the quantum simulation scheme proposed here can be implemented with existing spin control technology based on NV centers. The initialization, the pulses used to construct the operations, and the readout can be realized using a state of the art optically detected magnetic resonance equipment. In the realistic implementation of the scheme, the decoherence due to the unwanted electron spin-environment coupling is the major limitation of the simulation accuracy. One needs to accomplish the simulation as fast as possible compared to the electron spin decoherence time. NV centers are among the most promising spin qubits in solids, having quite long coherence times even at room temperature. In $^{13}$C natural diamond, the $T_{2e}^*$ and $T_{2e}$ of the NV electron spin are of the order of 3 $\mu s$ and 200 $\mu s$ respectively. To numerically estimate the operation time, the following numbers are used, which are feasible within current experimental technology: the duration of a strong microwave pulse is of the order of 0.05 $\mu s$, the duration of a $^{14}$N nuclear spin controlled selective microwave pulse is of the order of 0.5 $\mu s$ ($\sim 1/J_N$), the duration of RF pulses on $^{14}$N and $^{13}$C nuclear spin are of the order of 20 $\mu s$ and 2 $\mu s$ respectively, and the free evolution under the hyperfine coupling $J_C S_z C_z$ is of the order of 0.07 $\mu s$ ($\sim 1/J_C$). The operation time of each run of the procedure is estimated to be roughly of the order of 95 $\mu s$ (the number of slices in the Trotter approximation is chosen to be $n=2$, in which a minimum state fidelity of 0.87 has been obtained in our numerical computation compared to the evolution of the original Hamiltonian $H_S^B$). The operations on nuclear spins (RF pulses) contribute mainly to this time, which have a total time of the order of 88  $\mu s$. To reduce the electron spin decoherence for the duration of the RF pulses, one can insert a rapid electron spin flip in the middle and at the end of each RF pulse. Such an operation effectively eliminates the static electron spin-environment coupling and increases the decoherence time from $T_{2e}^*$ to $T_{2e}$ \cite{Huang11}. Therefore the effect of decoherence during nuclear spin operations is mild. The total time of electron spin operations is roughly 3 $\mu s$. This time is comparable to the $T_{2e}^*$. Hence the decoherence in these operations may have a considerable impact on the simulation accuracy. However, this problem can be solved by using isotopically-purified diamond. $T_{2e}^* \approx 200 \mu s$ and $T_{2e} \approx 3 ms$ have been reported in such diamonds with $^{12}$C abundance $>99.99\% $ \cite{Zhaonan}. At the same time, the reduction of the $^{13}$C concentration will also reduce the probability of finding a useful $^{13}$C nuclear spin in the vicinity of an NV center. Therefore the tradeoff between isotope concentration and spin qubit resources should be carefully considered. Adopting decoherence-protected quantum gate technologies \cite{Xu12,van12} may further reduce decoherence. Thus a high accuracy quantum simulation can be implemented.

In summary, we have devised a digital quantum simulation scheme based on NV center spins in diamond to simulate a QPT in ultrathin film TI. The scheme is feasible with current experimental techniques. This work shows the possibility of using NV centers as a promising quantum simulator to study important physical problems. We anticipate more physics in thin film TI \cite{Tse10,Seradjeh09,Yu10} can be simulated analogously using the proposed device. Other problems in TI may also be studied in this way, such as the design of a quantum circuit to extract the topological index directly (the $Z_2$ index for example \cite{Kane05a}), which does not possess a directly related physical quantity measurable in experiment. One can also simulate more realistic TI models of greater complexity than the simple Dirac model with resources that increase logarithmically with the number of bands compared to the linear increase seen in classical computers. The same advantages also appear in research on interacting TI \cite{Culcer11} and fractional TI \cite{Maciejko10,Levin09}.

This work was supported by the National Key Basic Research Program of China (Grant No. 2013CB921800), the National Natural Science Foundation of China (Grant Nos. 11227901, 91021005, 11104262), and the ¡¯Strategic Priority Research Program (B)¡¯ of the CAS (Grant No. XDB01030400).


\begin{references}
\bibitem{Gruber97} A. Gruber \textit{et al.}, Science \textbf{276}, 2012 (1997).

\bibitem{Fuchs09} G. D. Fuchs \textit{et al.}, Science \textbf{326}, 1520 (2009).
\bibitem{Neumann10} P. Neumann \textit{et al.}, Science \textbf{329}, 542 (2010).
\bibitem{van12} T. van der Sar \textit{et al.}, Nature \textbf{484}, 82 (2012).
\bibitem{Shi10} F. Shi \textit{et al.}, Phys. Rev. Lett. \textbf{105}, 040504 (2010).
\bibitem{Togan10} E. Togan \textit{et al.}, Nature \textbf{466}, 730 (2010).
\bibitem{Bernien13} H. Bernien \textit{et al.}, Nature \textbf{497}, 86 (2013).

\bibitem{Balasubramanian08} G. Balasubramanian \textit{et al.}, Nature \textbf{455}, 648 (2008).
\bibitem{Maze08} J. R. Maze \textit{et al.}, Nature \textbf{455}, 644 (2008).
\bibitem{Staudacher13} T. Staudacher \textit{et al.}, Science \textbf{339}, 561 (2013).
\bibitem{Kucsko13} G. Kucsko \textit{et al.}, Nature \textbf{500}, 54 (2013).
\bibitem{Neumann13} P. Neumann \textit{et al.}, Nano. Lett. \textbf{13}, 2738 (2013).

\bibitem{Xu12} X. Xu \textit{et al.}, Phys. Rev. Lett. \textbf{109}, 070502 (2012).
\bibitem{Balasubramanian09} G. Balasubramanian \textit{et al.}, Nature Materials \textbf{8}, 383 (2009).

\bibitem{Neumann08} P. Neumann \textit{et al.}, Science \textbf{320}, 1326 (2008).
\bibitem{Dutt07} M. V. Gurudev Dutt \textit{et al.}, Science \textbf{316}, 1312 (2007).
\bibitem{Fuchs11} G. D. Fuchs \textit{et al.}, Nature Physics \textbf{7}, 789 (2011).

\bibitem{Yao12} N. Y. Yao \textit{et al.}, Nature Communications \textbf{3}, 800 (2012).
\bibitem{Rabl10} P. Rabl \textit{et al.}, Nature Physics \textbf{6}, 602 (2010).

\bibitem{Lloyd96} S. Lloyd, Science \textbf{273}, 1073 (1996).
\bibitem{Buluta09} I. Buluta and F. Nori, Science \textbf{326}, 108 (2009).
\bibitem{Cai2013} J. Cai \textit{et al.}, Nature Physics \textbf{9}, 168 (2013).
\bibitem{Kane05} C. L. Kane and E. J. Mele, Phys. Rev. Lett. \textbf{95},226801 (2005).
\bibitem{Bernevig06} B. A. Bernevig \textit{et al.}, Science \textbf{314}, 1757 (2006).
\bibitem{Fu07} L. Fu \textit{et al.}, Phys. Rev. Lett. \textbf{98}, 106803 (2007).
\bibitem{Qi11} X.-L. Qi and S.-C. Zhang, Rev. Mod. Phys. \textbf{83}, 1057 (2011).
\bibitem{Qi09} X.-L. Qi \textit{et al.}, Science \textbf{323}, 1184 (2009).
\bibitem{Fu08} L. Fu and C. L. Kane, Phys. Rev. Lett. \textbf{100}, 096407 (2008).
\bibitem{Hasan10} M. Z. Hasan and C. L. Kane, Rev. Mod. Phys. \textbf{82}, 3045 (2010).
\bibitem{Zhang10} Y. Zhang \textit{et al.}, Nature Physics \textbf{6}, 584 (2010).
\bibitem{Jiang12} Y. Jiang \textit{et al.}, Phys. Rev. Lett. \textbf{108}, 016401 (2012).
\bibitem{Lu10} H.-Z. Lu \textit{et al.}, Phys. Rev. B \textbf{81}, 115407 (2010).
\bibitem{Liu13} W. E. Liu and D. Culcer, arXiv:1310.0075 (2013).
\bibitem{Tse10} W.-K. Tse and A. H. MacDonald, Phys. Rev. Lett. \textbf{105}, 057401 (2010).
\bibitem{Seradjeh09} B. Seradjeh \textit{et al.}, Phys. Rev. Lett. \textbf{103}, 066402 (2009).
\bibitem{Yu10} R. Yu \textit{et al.}, Science \textbf{329}, 61 (2010).
\bibitem{Zyuzin11} A. A. Zyuzin \textit{et al.}, Phys. Rev. B \textbf{83}, 245428 (2011).
\bibitem{Culcer12} D. Culcer, Physica E \textbf{44}, 860 (2012).
\bibitem{Mazza12} L. Mazza \textit{et al.}, New J. Phys. \textbf{14}, 015007 (2012).
\bibitem{Bermudez10} A. Bermudez \textit{et al.}, Phys. Rev. Lett. \textbf{105}, 190404 (2010).
\bibitem{Goldman10} N. Goldman \textit{et al.}, Phys. Rev. Lett. \textbf{105}, 255302 (2010).
\bibitem{Liu10} C.-X. Liu \textit{et al.}, Phys. Rev. B \textbf{82}, 045122 (2010).
\bibitem{Jiang09} L. Jiang \textit{et al.}, Science \textbf{326}, 267 (2009).
\bibitem{Abrams99} D. S. Abrams and S. Lloyd, Phys. Rev. Lett. \textbf{83}, 5162 (1999).
\bibitem{Steiner10} M. Steiner \textit{et al.}, Phys. Rev. B \textbf{81}, 035205 (2010).
\bibitem{Huang11} P. Huang \textit{et al.},Nature Communications \textbf{2}, 570 (2011).
\bibitem{Zhaonan} N. Zhao \textit{et al.}, Nature Nanotechnology \textbf{7}, 657 (2012).
\bibitem{Kane05a} C. L. Kane and E. J. Mele, Phys. Rev. Lett. \textbf{95}, 146802 (2005).
\bibitem{Culcer11} D. Culcer, Phys. Rev. B \textbf{84}, 235411 (2011).
\bibitem{Maciejko10} J. Maciejko \textit{et al.}, Phys. Rev. Lett. \textbf{105}, 246809 (2010).
\bibitem{Levin09} M. Levin and A. Stern, Phys. Rev. Lett. \textbf{103}, 196803 (2009).

\end{references}
\end{document}